\documentclass[12pt]{iopart}

\usepackage{amsfonts}
\usepackage{amssymb}
\usepackage{soul}
\usepackage{esint}
\usepackage{cite}
\usepackage{graphicx}
\usepackage{color}
\usepackage{mathrsfs}

\begin{document}

\title[Passive advection by random layered flows]{Passive advection of fractional Brownian motion by random layered flows}

\author{Alessio Squarcini$^{1,2}$, Enzo Marinari$^{3,4}$ \& Gleb Oshanin$^{5,6}$}

\address{
$^1$ Max-Planck-Institut f\"ur Intelligente Systeme, Heisenbergstr. 3, D-70569, Stuttgart, Germany \\
$^2$ IV. Institut f\"ur Theoretische Physik, Universit\"at Stuttgart, Pfaffenwaldring 57, D-70569 Stuttgart, Germany \\
$^3$ Dipartimento di Fisica, Sapienza Universit{\`a} di Roma, P.le A. Moro 2, I-00185 Roma, Italy \\
$^4$ INFN, Sezione di Roma 1 and Nanotech-CNR, UOS di Roma, P.le A. Moro 2, I-00185 Roma, Italy \\
$^5$  Sorbonne Universit\'e, CNRS, Laboratoire de Physique Th\'eorique de la Mati\`{e}re Condens\'ee (UMR 7600), 4 Place Jussieu, 75252 Paris Cedex 05, France\\
$^6$  Interdisciplinary Scientific Center J.-V. Poncelet (UMI CNRS 2615), Bolshoy
Vlasyevskiy Pereulok 11, 119002 Moscow, Russia
}

\ead{squarcio@is.mpg.de}
\vspace{10pt}
\begin{indented}
\item[]September 2019
\end{indented}

\begin{abstract}
\noindent We study statistical properties of the process $Y(t)$ of a passive advection by quenched 
random layered flows in situations when the inter-layer transfer is governed by a
 fractional Brownian motion $X(t)$ with the Hurst index $H \in (0,1)$.
We show that the disorder-averaged mean-squared displacement of the passive advection grows
in the large time $t$ limit in proportion to $t^{2 - H}$, which defines a family of anomalous super-diffusions. 
We evaluate the disorder-averaged Wigner-Ville spectrum of the advection process $Y(t)$ 
and demonstrate that it has a rather unusual 
power-law form $1/f^{3 - H}$ with a characteristic exponent which exceed the value $2$. Our results also
suggest that sample-to-sample fluctuations of the spectrum can be very important.
\end{abstract}


\maketitle


\section{Introduction}
The power spectral density of a time-dependent stochastic processes $Y(t)$  is a meaningful feature of its spectral content which describes how its power is distributed over frequency. For \textit{stationary} processes, it is usually defined as the time-average of the form
\begin{equation}
\label{spec1}
S(f) = \lim_{T \rightarrow \infty} \frac{1}{T} \mathbb{E}\left\{\left| \int^T_0 \textrm{d}t\; \, \textrm{e}^{i f t}\; Y(t)\right|^2\right\} \,,
\end{equation}
where the symbol $\mathbb{E}\left\{\ldots\right\}$ here and henceforth denotes the expected value with respect to different realizations of the process $Y(t)$. For \emph{non-stationary} processes, however, the expression given in (\ref{spec1}) may not make sense and even the limit $T \to \infty$ on the right hand side of it may not exist, so that one has to seek other meaningful interpretations of the power spectral density (see e.g., Refs. \cite{man,loynes,flandrin2,flandrin1,lutz,Krapf_2018,Krapf_2019,spos}). Although no unique tool exists for performing a time-dependent spectral analysis of non-stationary random functions, one of the physically plausible approaches \cite{flandrin2,flandrin1} consists in using the time-averaged functional 
\begin{equation}
\label{spec2}
W_{f} = \lim_{T \to \infty} \frac{1}{T} \int^{T}_0 \textrm{d}t \, W(f, t) \,,
\end{equation}
where $W(f, t)$ is the Wigner-Ville spectrum
\begin{equation}
\label{spec3}
W(f, t) = \int^{\infty}_{-\infty} \textrm{d}\tau \textrm{e}^{- i f \tau} \, \mathbb{E}\left\{Y\left(t+\frac{\tau}{2}\right) Y\left(t - \frac{\tau}{2}\right)\right\} \,.
\end{equation}
The functions $S_f$ in (\ref{spec1}) and $W_f$ in (\ref{spec2}) and (\ref{spec3}) become identical, once $Y(t)$ is stationary.

Many naturally occurring processes, as well as many processes encountered in engineering and technological sciences, exhibit power spectra of the form $\sim A/f^{\alpha}$, where $A$ is an $f$-independent amplitude. The characteristic exponent $\alpha$ may be as small as $\alpha = 1$ in the case of flicker noise \cite{dutta} or standard Sinai diffusion \cite{enzo,enzo1}, $\alpha = 2$ in the paradigmatic case of Brownian motion \cite{dutta,Krapf_2018}\footnote{Note that while
$\alpha = 2$ is a valid result for Brownian motion, the observation of the $A/f^2$ law alone does not imply that one necessarily deals with the standard diffusion. The $A/f^2$ law holds for the finite-$T$ 
power spectra in (\ref{spec1}) of super-diffusive fractional Brownian motion with the Hurst index $H > 1/2$ \cite{Krapf_2018}, anomalous scaled diffusion \cite{spos}, a variety of the so-called diffusing-diffusivity models \cite{spos1} and also 
for the running maximum of Brownian motion \cite{gleb1} and diffusion in periodic Sinai chains \cite{gleb2}, to name but a few processes. The amplitude $A$ in the first three examples is, however, ageing, i.e., it is dependent on the observation time $T$.} and may even exceed the value of $2$, e.g., for the Wigner-Ville power spectra in (\ref{spec2}) and (\ref{spec3}) of a super-diffusive fractional Brownian motion, for which one
 has $\alpha = 2 H + 1$ \cite{flandrin1}. While the examples with $\alpha=1$ or $\alpha > 2$ are rather rare, there exist numerous processes for which the power spectrum is a power-law function with  $1 < \alpha < 2$.  Few stray examples include electrical signals in vacuum tubes, semiconductor devices and metal films \cite{man,dutta}. More generally, such a behaviour is observed in sequences of earthquakes \cite{5} and weather data \cite{wea}, in evolution \cite{7}, human cognition \cite{8}, network traffic \cite{9}, fractional Brownian motion with stochastic reset \cite{maj} and even in the distribution of loudness in musical recordings \cite{press}.  Recent experiments have also revealed such a behaviour of spectra for transport in individual ionic channels \cite{10,11}, electrochemical signals in nanoscale electrodes \cite{13}, bio-recognition processes \cite{14} and intermittent quantum dots \cite{lutz,15}.  Many other examples and unresolved problems have been discussed in Refs. \cite{lutz,15,mike,16,17}.

In this paper we discuss a physical model with a super-diffusive dynamics and calculate its Wigner-Ville power spectral density defined in (\ref{spec2}) and (\ref{spec3}),  which is shown here to exhibit a power-law with an exponent $\alpha$ exceeding $2$. We focus on a stochastic passive advection process $Y(t)$ in the presence of quenched, random layered flows. The model has been introduced originally by Dreizin and Dykhne \cite{dykhne} for the analysis of conductivity of inhomogeneous media in a strong magnetic field, and by Matheron and de Marsily \cite{MdM} for the analysis of transport of solute in a stratified porous medium with flow parallel to the bedding. Subsequently, different facets of this model have been analysed in a great detail (see, e.g., Refs. \cite{redner,ledoussal1,ledoussal2,blumen1,ledoussal3,blumen2,crisanti,majumdar,roy}), establishing also a link, on a mathematical level, to the well-known models of statistical mechanics such as diffusion in presence of sources and sinks, spin depolarisation in random fields, self-repulsive polymers, and an electron in a random potential (see Ref. \cite{ledoussal3}). The model has been also generalised to study dynamics of more complicated objects, e.g., flexible polymers, subject to such velocity fields (see, e.g., Refs. \cite{blumen3,ledoussal4,jespersen,majumdar2,simi,katyal}). While in the original settings in Refs. \cite{dykhne,MdM} a random motion $X(t)$ in the direction  perpendicular to the layered velocity fields was supposed to be diffusive,  these latter works \cite{blumen3,ledoussal4,jespersen,majumdar2,simi,katyal} provided some insight into the statistical properties of the process $Y(t)$ for \textit{anomalous} inter-layer diffusion. In particular, when $X(t)$ is a standard diffusion, one finds the super-diffusive behaviour of the form 
$\langle \mathbb{E}\left\{Y^2(t)\right\}\rangle \sim t^{3/2}$ for the displacement along the direction of the flows, where the angle brackets here and henceforth denote averaging with respect to the distribution of the flow velocities. For a tagged bead of an infinite Rouse polymer, an even stronger super-diffusion has been predicted \cite{blumen3}, $\langle \mathbb{E}\left\{Y^2(t)\right\}\rangle \sim t^{7/4}$. Here, we consider a more general case when $X(t)$ is a fractional Brownian motion with the Hurst exponent $H$, $(0 < H < 1)$, and derive an $H$-parametrised family of super-diffusive laws for $\langle \mathbb{E}\left\{Y^2(t)\right\}\rangle$ and the corresponding family of the Wigner-Ville power spectra in (\ref{spec2}), (\ref{spec3}) of the form $\langle W_f \rangle = A/f^{\alpha}$ with $\alpha > 2$.

The paper is outlined as follows: In Sec. \ref{mod} we describe our model and introduce basic notations. In Sec. \ref{dis1}, we calculate the disorder-averaged mean-squared displacement along the $Y$-axis and also quantify
 its sample-to-sample fluctuations by analysing the coefficient $\gamma_v$ of variation of the corresponding distribution of $\mathbb{E}\left\{Y^2(t)\right\}$. We show that 
 for the Hurst index $H \gtrsim 0.22$, the coefficient $\gamma_v$ of variation is less than unity, meaning that the sample-to-sample fluctuation are significant but their overall effect is not dramatic. On contrary, for $0< H \lesssim 0.22$, $\gamma_v$ exceeds unity which implies that the standard deviation becomes larger than the mean value. Therefore, for such $H$, one expects the sample-to-sample variations of  $\mathbb{E}\left\{Y^2(t)\right\}$ to become much more important. Further on, 
 in Sec. \ref{dis2} we turn to the central point of our analysis - calculation of 
 the disorder-averaged Wigner-Ville spectral density  $W_f$ of the passive advection process. We show that the latter exhibits the frequency $f$-dependence of the form $A/f^{3 - H}$, i.e., has an exponent $\alpha$ which exceeds the value $2$. Here we also attempt to quantify the effective broadness of the distribution of the realisation-dependent value of  $W_f$. To this end, we consider the behaviour of $W_f$ for a finite observation time $T$ and zero frequency, $f = 0$.
 We show that in this particular case, the coefficient of variation of the distribution grows (as a power law) with $T$, which signals that the sample-to-sample fluctuations are most important in this limit. Next, in Sec. \ref{dis3} we concentrate on the representation of the Wigner-Ville spectrum in form of an Ising-type chain of "spins" $\sigma_k$ and analyse the behaviour of the effective time-dependent couplings $J_{k,k'}(t)$ between the "spins" $\sigma_k$ and $\sigma_{k'}$. Finally, we conclude in Sec. \ref{dis4} with a brief recapitulation of our results and outline the perspectives of further research.
 
\section{The model}
\label{mod}
\begin{figure}[t]
\begin{center}
\centerline{\includegraphics[width = 110mm]{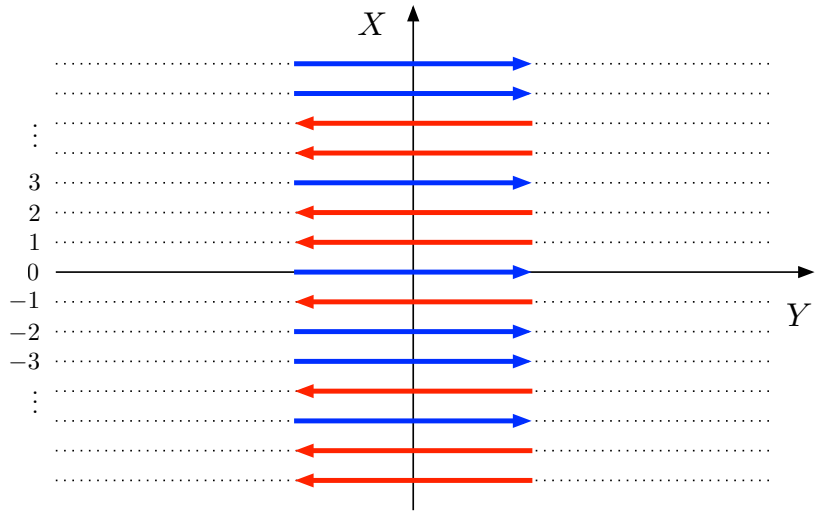}}
\caption{A realisation of a pattern of random layered flows. The particle undergoes a fractional Brownian motion along the $X$-direction and is passively advected along the quenched random flows in the $Y$-direction.
\label{Fig1}}
\end{center}
\end{figure}

Consider the dynamics of a particle in the two-dimensional model system depicted in Fig.\ref{Fig1}, which we define as follows: a particle undergoes an unbiased random motion $X(t)$, (starting at the origin at $t=0$, $X(0)=0$), along the $X$-axis. This random motion is the so-called fractional Brownian motion (fBm) \cite{ness}. The latter is formally defined as a stochastic integral with respect to the white noise measure $\textrm{d}B_s$:
\begin{equation}
\label{ }
\fl X(t) = \frac{1}{\Gamma(H+1/2)}\biggl[ \int_{-\infty}^{0}\textrm{d}B_{s} \left( (t-s)^{H-1/2} - (-s)^{H-1/2} \right) + \int_{0}^{t}\textrm{d}B_{s} (t-s)^{H-1/2} \biggr]
\end{equation}
where $H \in (0,1)$ is called the Hurst index. This is a Gaussian process with zero mean and 
covariance 
\begin{equation}
\label{cov}
\mathbb{E}\left\{X(t_1) X(t_2)\right\} = \frac{1}{2} \left(t_1^{2 H} + t_2^{2 H} - |t_1 - t_2|^{2 H}\right) \,.
\end{equation}
The standard Brownian motion with independent increments is recovered for $H = 1/2$. When $H \neq 1/2$, the increments are correlated so that for $H > 1/2$ if there is an increasing pattern in the previous steps, then it is likely that the current step will be increasing as well, resulting ultimately in a super-diffusive motion. For $H < 1/2$ the increments are negatively correlated, which entails a sub-diffusive motion.

Further on, we mark along the $X$-axis the points $X_k = x \, k$, $k=0, \pm 1, \pm 2, \ldots$ (dashed lines in Fig. \ref{Fig1}), where $x$ is the distance between adjacent points. For simplicity, we set this distance equal to unity. At these points we have flows, depicted by arrows  in Fig.\ref{Fig1}, with a velocity which is constant along the $Y$-axis (this constant is set equal to $1$) and with $k$-dependent orientation described by a quenched random variable   $\sigma_k = \pm 1$, such that $\sigma_k = + 1$ corresponds to a flow in the positive $Y$-direction (positive velocity), and $\sigma_k = - 1$ corresponds to a flow orientated in the negative $Y$-direction. We concentrate here solely on the case when $\sigma$-s are delta-correlated and when there is no global flow in the system, i.e., when
$\langle \sigma_k\rangle \equiv 0$.

Next, we assume that when the particle appears in-between the neighbouring points $X_k$, it does not experience any drift in the $Y$-direction. However, once it arrives at any $X_k$, it is passively (and instantaneously) advected on a fixed distance $y = 1$ along the direction defined by the arrow (flow) at this point. Consequently, the current position of the particle along the $Y$-axis obeys
\begin{equation}
\label{Y}
Y(t) = \sum_{k=-\infty}^{\infty} {\cal N}_k[X(t)] \, \sigma_k \,,
\end{equation}
where ${\cal N}_k[X(t)]$ is a trajectory $X(t)$-dependent random variable - the so-called local time \cite{local1,local2} - which measures how many times within the time interval $(0,t)$ the point $X_k$ has been visited by $X(t)$. This random variable can be formally represented as
\begin{equation}
\label{def}
{\cal N}_k[X(t)] = \int^t_0 \textrm{d}\tau \, \delta\left(k - X(\tau)\right) \,,
\end{equation}
where $\delta(\ldots)$ denotes the Dirac delta-function.

\section{Disorder-averaged mean-squared displacement of $Y(t)$.}
\label{dis1}
We analyse first the mean-squared displacement of the particle along the $Y$-axis. Thanks to (\ref{Y}), we can formally write this average as
\begin{equation}
\label{mean}
\mathbb{E}\left\{Y^2(t)\right\} = \sum_{k=-\infty}^{\infty} \sum_{k'=-\infty}^{\infty} \sigma_k \sigma_{k'} \mathbb{E}\left\{{\cal N}_k[X(t)] {\cal N}_{k'}[X(t)]\right\} \, .
\end{equation}
In order to calculate the expected value of the product of two local times taken at two different positions in space, we need to know the two-point probability distribution function for the fBm. The probability $P(X(\tau_1)|\tau_1;X(\tau_2)|\tau_2;0|0)$ that a fBm, starting at the origin, appears at position $X(\tau_1)$ at time moment $\tau_1$ and at position $X(\tau_2)$ at time moment $\tau_2$ ($\tau_1$ and $\tau_2$ being unordered) is given by
\begin{eqnarray}
\label{2point}
\fl P(X(\tau_1)|\tau_1;X(\tau_2)|\tau_2;0|0)  &=&  \frac{1}{2 \pi \tau_1^H \tau_2^H \sqrt{1 - g^2}} \nonumber \\ \nonumber
\fl & \times & \exp\Biggl[- \frac{1}{2 (1-g^2)} \left(\frac{X^2(\tau_1)}{\tau_1^{2 H}} + \frac{X^2(\tau_2)}{\tau_2^{2 H}} - 2 g \frac{X(\tau_1) X(\tau_2)}{\tau_1^H \tau_2^H}\right)\Biggr] \, , \\
\end{eqnarray}
where $g$ is the correlation coefficient
\begin{equation}
\label{def_g}
g = \frac{\mathbb{E}\left\{X(t_1) X(t_2)\right\}}{\tau_1^H \tau_2^H} = \frac{\tau_1^{2 H} + \tau_2^{2 H} - |\tau_1 - \tau_2|^{2 H}}{2 \tau_1^H \tau_2^H} \,.
\end{equation}
Since $|\tau_1^H - \tau_2^H| \leq |\tau_1 - \tau_2|^H$, we have that $0 \leq g \leq 1$. Therefore, by using (\ref{2point}) we readily find that
\begin{eqnarray}
\label{zy}
\mathbb{E}\left\{{\cal N}_k[X(t)] {\cal N}_{k'}[X(t)]\right\} & = & \int_{0}^{t}\textrm{d}\tau_{1}\int_{0}^{t}\textrm{d}\tau_{2} \, P(k \vert \tau_{1}, k^{\prime} \vert \tau_{2} , 0\vert0) \, .
\end{eqnarray}
The above expression represents the desired two-point correlation function of the local times of a fBm at two different (or coinciding) points taken at the same time moment $t$. By inserting (\ref{zy}) into (\ref{mean}), performing the averaging over the distribution of $\{\sigma_k\}$ and summation over $k$, (as well as appropriately changing the integration variables), we get
\begin{eqnarray} \nonumber
\langle \mathbb{E}\left\{Y^2(t)\right\} \rangle & = & \sum_{k=-\infty}^{\infty} \mathbb{E}\left\{{\cal N}^2_k[X(t)]\right\} \\ \nonumber
& = & \frac{t^{2 - 2 H}}{2 \pi} \int^1_0 \frac{{\rm d}x_1}{x_1^{H}} \int^1_0 \frac{{\rm d}x_2}{x_2^H \sqrt{1 - \phi^2(x_1,x_2)}} \times \\ 
& \times & \theta_3\left(0, \exp\left(- \frac{|x_1-x_2|^{2 H}}{2 t^{2 H} \left(1 - \phi^2(x_1,x_2)\right) x_1^{2 H} x_2^{2 H}}\right)\right) \,,
\end{eqnarray}
where 
\begin{equation}
\label{phi}
\phi(x_1,x_2) = \frac{x_1^{2 H} + x_2^{2 H} - |x_1 - x_2|^{2 H}}{2 x_1^H x_2^H} \leq 1 \,,
\end{equation}
and $\theta_3(\ldots)$ is the Jacobi's theta function. Turning  to the limit $t \to \infty$, we find
\begin{eqnarray} \nonumber
\label{m}
& & \theta_3\left(0, \exp\left(- \frac{|x_1-x_2|^{2 H}}{2 t^{2 H} \left(1 - \phi^2(x_1,x_2\right) x_1^{2 H} x_2^{2 H}}\right)\right) \\
& = &\sqrt{2 \pi} \frac{\sqrt{1 - \phi^2(x_1,x_2)} x_1^H x_2^H}{|x_1-x_2|^H} t^H + o\left(t^H\right) \,,
\end{eqnarray}
where the symbol $o\left(t^H\right)$ means that the omitted terms grow with $t$ slower than $t^H$. 
Note that taking into account only the leading in the limit $t \to \infty$ term in the right hand side of (\ref{m}), is tantamount to converting the summation over $k$ into an integral - by using the Euler-Maclaurin summation formula - and discarding all the correction terms. Consequently, we arrive at the following asymptotic large-$t$ form 
\begin{eqnarray} \nonumber
\langle \mathbb{E}\left\{Y^2(t)\right\}\rangle & = & \frac{t^{2 - H}}{\sqrt{2 \pi}} \int^1_0 \int^1_0 \frac{{\rm d}x_1 {\rm d}x_2}{|x_1-x_2|^H}  + o\left(t^{2-H}\right) \\ 
& = & \sqrt{\frac{2}{\pi}} \frac{t^{2 - H}}{(1-H)(2 - H)} + o\left(t^{2-H}\right)
\end{eqnarray}
which represents the desired result on the $H$-parametrised family of anomalous laws describing the passive advection of a fBm by random layered flows. Note that for any $H \in (0,1)$ the disorder-averaged mean-square displacement exhibits a super-diffusive behaviour. In particular, for $H = 1/2$ we recover the result $\langle \mathbb{E}\left\{Y^2(t)\right\}\rangle \sim t^{3/2}$ obtained previously in Refs. \cite{dykhne,MdM}.  For $H = 1/4$, we find an even stronger super-diffusive law 
 $\langle \mathbb{E}\left\{Y^2(t)\right\}\rangle \sim t^{7/4}$, obtained earlier in Ref. \cite{blumen3} for the dynamics of a tagged monomer in an infinite Rouse polymer in the presence of such random flows. In general, we observe that the smaller $H$ is, the more pronounced becomes the super-diffusive behaviour of $Y(t)$, which is a simple consequence of the fact that for a more spatially confined sub-diffusion, the particle spends more time in a given layer, and hence,  is advected on larger scales.

\subsection{Sample-to-sample fluctuations of the mean-squared displacement.}
\begin{figure}[htbp]
\begin{center}
\centerline{\includegraphics[width = .65 \textwidth]{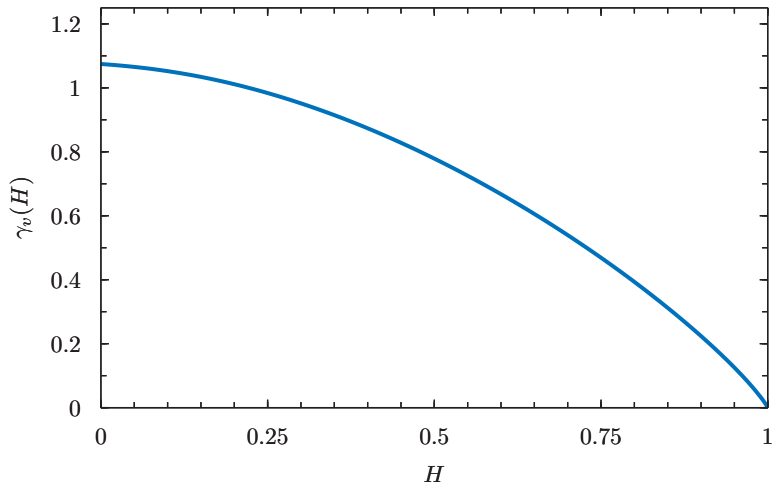}}
\caption{The coefficient $\gamma_v$ of variation of the distribution function of the random variable $\mathbb{E}\left\{Y^2(t)\right\}$ as a function of $H$.
\label{Fig2}}
\end{center}
\end{figure}

Here we address the question of the sample-to-sample fluctuations of $\mathbb{E}\left\{Y^2(t)\right\}$ as defined in (\ref{mean}) for different realisations of the patterns of random flows. In order to quantify these fluctuations, we focus on the \textit{variance} of  $\mathbb{E}\left\{Y^2(t)\right\}$, which is defined as follows
\begin{eqnarray} \nonumber
\label{varvar}
{\rm Var}_{\sigma}\left(\mathbb{E}\left\{Y^2(t)\right\}\right) & = & \langle \mathbb{E}^2\left\{Y^2(t)\right\} \rangle -  \langle \mathbb{E}\left\{Y^2(t)\right\} \rangle^2 \\
& = & 2 \sum_{k=-\infty}^{\infty} \sum_{k' = - \infty, k' \neq k}^{\infty}  \mathbb{E}^2\left\{{\cal N}_k[X(t)] {\cal N}_{k'}[X(t)]\right\} \, .
\end{eqnarray}
We turn to the limit $t \to \infty$ and concentrate on the leading term which dominates the large-time asymptotic behaviour of the variance. It is plausible then to convert the infinite sums in (\ref{varvar}) into integrals, which can be readily made dimensionless by an appropriate change of the integration variables. In doing so, we get
\begin{equation}
{\rm Var}_{\sigma}\left(\mathbb{E}\left\{Y^2(t)\right\}\right) = \frac{2 \gamma_{v}^2}{\pi (1-H)^2 (2 - H)^2} \, t^{4 - 2 H} \,,
\end{equation}
with $\gamma_v$ being the (time-independent) 
coefficient of variation of the distribution $P_{\sigma}\left(\mathbb{E}\left\{Y^2(t)\right\}\right)$ of $\mathbb{E}\left\{Y^2(t)\right\}$. This coefficient is defined by
\begin{eqnarray}
\gamma_v & = & \sqrt{\frac{{\rm Var}_{\sigma}\left(\mathbb{E}\left\{Y^2(t)\right\}\right)}{\langle \mathbb{E}\left\{Y^2(t)\right\}\rangle^2}}  \,.
\end{eqnarray}
A straightforward calculation allows us to write the following expression for $\gamma_v $:
\begin{equation}
\label{gamma_integral}
\gamma_v = \frac{(1 - H)(2 - H)}{\sqrt{2}} \left(\int^1_0 \int^1_0 \int^1_0 \int^1_0 \frac{{\rm d}x_1 {\rm d}x_2 {\rm d}x_3 {\rm d}x_4}{\sqrt{q}}\right)^{1/2} \, ,
\end{equation}
where $q$, for a given $H$, is a numerical factor defined by
\begin{eqnarray} \nonumber
q & = & (x_1 x_2)^{2 H} + (x_2 x_3)^{2 H} + (x_3 x_4)^{2 H} + (x_4 x_1)^{2 H} \\
& - & \frac{1}{4} \left(x_1^{2 H} + x_2^{2 H} +x_3^{2 H} + x_4^{2 H} - |x_1 - x_2|^{2 H} - |x_3 - x_4|^{2 H}\right)^2 \, .
\end{eqnarray}
The coefficient of variation $\gamma_{v}$ is a meaningful characteristic of the effective broadness of the distribution showing how $\mathbb{E}\left\{Y^2(t)\right\}$ fluctuates from sample to sample. Being unable to perform exactly 
the four-fold integral (\ref{gamma_integral}) entering the expression for $\gamma_v$, we evaluate it numerically and plot it as a function of the Hurst index in Fig. \ref{Fig2}.
We observe that $\gamma_v$ is a monotonically decreasing function of $H$ which perfectly makes sense because for a fixed time interval $(0,t)$ the span of the trajectory $X(t)$ is a monotonically increasing function of $H$. Further on, we notice that $\gamma_v(H)$ is less than unity for most of the values of $H$ in the interval $(0,1)$ meaning that sample-to-sample fluctuations here are not very significant. The coefficient of variation exceeds slightly $1$ (such that the standard deviation becomes greater than the mean value) only for $H \lesssim 0.22$; note that $\gamma_v(0)=(4/3)^{1/4} \approx 1.075$. Only in this region, i.e., for very spatially confined sub-diffusive processes $X(t)$, the sample-to-sample fluctuations may become rather significant (see, e.g., Ref. \cite{carlos,carlos2}).

\section{Disorder-averaged Wigner-Ville power spectral density $W_f$.}
\label{dis2}
We turn next to the calculation of the corresponding power spectral density of the passive advection process $Y(t)$. To this end, it is first expedient to somewhat simplify these expressions. We formally rewrite the covariance of the process $Y(t)$ as 
\begin{equation}
\mathbb{E}\left\{Y(t_1) Y(t_2)\right\} = \int^{t_1}_0 {\rm d}\tau_1 \int^{t_2}_0 {\rm d}\tau_2 \,\, Q(\tau_1,\tau_2) \,,
\end{equation}
with
\begin{equation}
\label{ }
Q(\tau_1,\tau_2) = \sum_{k=-\infty}^{\infty} \sum_{k'=-\infty}^{\infty} \sigma_k \sigma_{k'}  \,\, \mathbb{E}\left\{\delta\left(k - X(\tau_1)\right) \delta\left(k' - X(\tau_2)\right) \right\} \, .
\end{equation}
The Wigner-Ville spectrum of the process $Y(t)$ then reads
\begin{eqnarray}
\label{1}\nonumber
\fl W(f,t) & = \int^{\infty}_{-\infty} {\rm{d}}\tau {\rm{e}}^{- i f \tau} \int^{t+\tau/2}_0 {\rm{d}}\tau_1 \int^{t-\tau/2}_0 {\rm{d}}\tau_2 \, Q(\tau_1,\tau_2) \\
\fl & = \int^{\infty}_{-\infty} {\rm{d}}\tau {\rm{e}}^{- i f \tau} \int^{\infty}_0 {\rm{d}}\tau_1 \int^{\infty}_0 {\rm{d}}\tau_2 \, \theta\left(t + \frac{\tau}{2} - \tau_1\right) \, \theta\left(t - \frac{\tau}{2} - \tau_2\right) \, Q(\tau_1,\tau_2)
\end{eqnarray}
where $\theta(\ldots)$ denotes the Heaviside theta-function. In (\ref{1}), we can straightforwardly perform the integral over $\tau$ to get
\begin{equation}
\fl W(f,t) = - \frac{i}{f} \int^{\infty}_0 {\rm{d}}\tau_1 \int^{\infty}_0 {\rm{d}}\tau_2 \, \theta\left( t - \frac{\tau_1 + \tau_2}{2}\right) \, \left({\rm{e}}^{2 i f (t - \tau_1)} - {\rm{e}}^{- 2 i f (t - \tau_2)}\right) \, Q(\tau_1,\tau_2) \, .
\end{equation}
Inserting the latter expression into (\ref{spec3}) and performing the integration over $t$, we find 
\begin{eqnarray} \nonumber
\label{2}
W_f & = & - \frac{1}{2 f^2} \lim_{T \to \infty} \frac{1}{T} \int^{\infty}_0 {\rm{d}}\tau_1 \int^{\infty}_0 {\rm{d}}\tau_2 \, \theta\left(T- \frac{\tau_1 + \tau_2}{2}\right) \times \\
& \times & \left({\rm{e}}^{2 i f (T - \tau_1)} - {\rm{e}}^{- 2 i f (T - \tau_2)}\right)^2 \, Q(\tau_1,\tau_2) \, .
\end{eqnarray}
The expression in the right-hand-side of (\ref{2}) is still quite complicated since it involves oscillating functions of $\tau_1$ and $\tau_2$ and the integration limits are mixed. We notice then that the existence of the limit in the right-hand-side of (\ref{2}) implies that the two-fold integral  
\begin{equation}
\label{3}
 \int^{\infty}_0 {\rm{d}}\tau_1 \int^{\infty}_0 {\rm{d}}\tau_2 \, \theta\left(T- \frac{\tau_1 + \tau_2}{2}\right)  \, 
\left({\rm{e}}^{i f (T - \tau_1)} - {\rm{e}}^{- i f (T - \tau_2)}\right)^2 \, Q(\tau_1,\tau_2) 
\end{equation}
grows linearly with $T$ when $T \to \infty$. In turn, this means that the Laplace transform of the expression in (\ref{3}), (with respect to $T$ with the Laplace parameter $\lambda > 0$), diverges as $1/\lambda^2$ in the limit $\lambda \to 0$. This permits us to formally rewrite (\ref{2}) as 
\begin{equation}
\label{4}
W_f = 4 \lim_{\lambda \to 0} \frac{\lambda}{4 f^2 + \lambda^2} \int^{\infty}_0 {\rm d} \tau_1 \int^{\infty}_0 {\rm d}\tau_2 \, {\rm e}^{- i f(\tau_1 - \tau_2) - \frac{\lambda}{2}(\tau_1+\tau_2)} \, Q(\tau_1,\tau_2) \,,
\end{equation}
which appears to be more convenient for further calculations.

We focus next on the disorder-averaged Wigner-Ville spectrum $\langle W_f \rangle$. The disorder-averaged function $Q(\tau_1,\tau_2)$ reads
\begin{equation}
\label{5}
\langle Q(\tau_1,\tau_2) \rangle = \frac{1}{2 \pi \tau_1^H \tau_2^H \sqrt{1 - g^2}} \theta_3\left(0, \exp\left(- \frac{|\tau_1 - \tau_2|^{2 H}}{2 \tau_1^{2 H} \tau_2^{2 H} \left(1 - g^2\right)}\right)\right) \,.
\end{equation}
Since we are interested 
the behaviour 
in the limit $\lambda \to 0$,  
we have to focus on the asymptotic behaviour of the Jacobi theta-function in the limit $\tau_1 \to \infty$ and $\tau_2 \to \infty$. In this limit the expression in the exponential in the theta-function tends to zero such that we find that, in the leading order, the theta-function behaves as
\begin{equation}
\theta_3\left(0, \exp\left(- \frac{|\tau_1 - \tau_2|^{2 H}}{2 \tau_1^{2 H} \tau_2^{2 H} \left(1 - g^2\right)}\right)\right)  \approx \sqrt{2 \pi \left(1 - g^2\right)} \left(\frac{\tau_1 \tau_2}{|\tau_1 - \tau_2|}\right)^H \,,
 \end{equation}
which yields
\begin{equation}
\label{Q}
\langle Q(\tau_1,\tau_2) \rangle \approx \frac{1}{\sqrt{2 \pi} |\tau_1 - \tau_2|^H} \,.
\end{equation}
Inserting this expression into the two-fold integral in (\ref{4}) and performing the integrations, we find
\begin{equation}
\fl \int^{\infty}_0 {\rm d}\tau_1 \int^{\infty}_0 {\rm d}\tau_2 \, {\rm e}^{- i f(\tau_1 - \tau_2) - \frac{\lambda}{2}(\tau_1+\tau_2)} \, \langle Q(\tau_1,\tau_2) \rangle = \sqrt{\frac{2}{\pi}} \frac{\Gamma(1-H) \, \sin\left(\frac{\pi H}{2}\right)}{f^{1-H}} \frac{1}{\lambda} + O\left(1\right) \,,
\end{equation} 
where the symbol $O\left(1\right)$ means that the omitted terms are constant in the limit $\lambda \to 0$. 
Inserting the latter expression into (\ref{4}) and taking subsequently the limit $\lambda \to 0$, we obtain the following result for the disorder-averaged 
Wigner-Ville power spectral density:
\begin{equation}
\label{dis-av}
\langle W_f \rangle = \sqrt{\frac{2}{\pi}} \frac{\Gamma(1-H) \sin\left(\frac{\pi H}{2}\right)}{f^{3 - H}} \,.
\end{equation}
This result is the central point of our analysis and
defines an $H$-parametrised family of super-diffusive spectra characterised by the exponent $\alpha=3-H > 2$ for any $0<H<1$. Notice also that the $H$-dependent overall factor is a positive definite, monotonically increasing function of $H$. We further note that for the case when $X(t)$ is a standard Brownian motion, we have
\begin{equation}
\label{BMS}
\langle W_f \rangle = \frac{1}{f^{5/2}} \,,
\end{equation}
while for $H = 1/4$ we get
 \begin{equation}
\langle W_f \rangle = \frac{\mathscr{C}}{f^{11/4}} \,, \qquad \mathscr{C} = \sqrt{\frac{2-\sqrt{2}}{2\pi}}\Gamma(3/4) \simeq 0.374 \, ,
\end{equation}
which particular 
case corresponds to the dynamics of a tagged bead in an infinite Rouse chain.

\subsection{Sample-to-sample fluctuations of the Wigner-Ville spectrum at zero-frequency}

Consider the following finite-time averaged Wigner-Ville spectrum in (\ref{spec2}) and (\ref{spec3}) at a finite observation time $T$ (such that we drop the limit $T \to \infty$ in (\ref{spec2})) and denote the resulting expression  as $W_{f}^{(T)}$. 
Then, the disorder-averaged $W_{f}^{(T)}$ reads
\begin{equation}
\label{zf1}
\fl \langle W_{f}^{(T)} \rangle = - \frac{1}{2f^{2}T} \int_{0}^{\infty}\textrm{d}\tau_{1}\int_{0}^{\infty}\textrm{d}\tau_{2} \, \theta(2T-\tau_{1}-\tau_{2}) \left( \textrm{e}^{if(T-\tau_{1})} - \textrm{e}^{-if(T-\tau_{2})} \right)^{2} \langle Q(\tau_{1},\tau_{2}) \rangle \, ,
\end{equation}
with $\langle Q(\tau_{1},\tau_{2}) \rangle$ given by (\ref{Q}).
In the limit $f \to 0$ the integrand in the latter expression stays finite. 
 Performing next the change of the integration variables $\tau_{1}=T x_{1}$, $\tau_{2}=T x_{2}$, we realise that the asymptotic behaviour of the finite-time disorder-average Wigner-Ville spectrum follows
\begin{equation}
\label{mean_zf}
\langle W_{0}^{(T)} \rangle \approx T^{1-H} \, \Psi_{1}(H) \, ,
\end{equation}
where $\Psi_{1}(H)$ is the function
\begin{eqnarray}\nonumber
\Psi_{1}(H) & = & \frac{1}{\sqrt{8\pi}} \int_{0}^{\infty}{\rm d}x_{1}\int_{0}^{\infty}{\rm d}x_{2} \frac{ \theta(2-x_{1}-x_{2}) (2-x_{1}-x_{2})^{2} }{ |x_{1}-x_{2}|^{H} } \\
& = & \frac{2^{-H+7/2}}{\sqrt{\pi}(1-H)(2-H)(3-H)(4-H)} \, .
\end{eqnarray}
Notice that for any $0<H<1$ the disorder-averaged $\langle W_{0}^{(T)} \rangle$ grows as a power law of the observation time $T$. For $H\rightarrow 1^{-}$ the exponent $1-H$ in the observation time vanishes while the overall factor $\Psi(H)$ diverges.

Next, in order to define the coefficient of variation in this zero-frequency limit, we have to evaluate 
the variance of the finite-time Wigner-Ville spectrum:
\begin{equation}
\label{var2}
\textrm{Var}\left(  W_{f}^{(T)} \right) = \biggl\langle \left( W_{f}^{(T)} \right)^{2} \biggr\rangle -  \biggl\langle W_{f}^{(T)} \biggr\rangle^{2} \, .
\end{equation}
To this end, introducing an auxiliary function
\begin{equation}
\label{ }
g(\tau_{1},\tau_{2},T) = \theta(2T-\tau_{1}-\tau_{2}) \left( \textrm{e}^{if(T-\tau_{1})} - \textrm{e}^{-if(T-\tau_{2})} \right)^{2} \, ,
\end{equation}
we formally express the variance as 
\begin{equation}
\label{var3}
\fl \textrm{Var}\left(  W_{f}^{(T)} \right) = \frac{1}{4f^{4}T^{2}} \int_{0}^{\infty}\textrm{d}\tau_{1} \cdots \int_{0}^{\infty}\textrm{d}\tau_{4} \, g(\tau_{1},\tau_{2},T) g(\tau_{3},\tau_{4},T) R(\tau_{1},\dots,\tau_{4})
\end{equation}
where
\begin{eqnarray} \nonumber
\label{functionR}
R(\tau_{1},\dots,\tau_{4}) & = & \sum_{k} P(k \vert \tau_{1}, k \vert \tau_{2}, 0\vert0)P(k \vert \tau_{3}, k \vert \tau_{4}, 0\vert0) + \\ \nonumber
& + & \sum_{k,n} P(k \vert \tau_{1}, n \vert \tau_{2}, 0\vert0)P(k \vert \tau_{3}, n \vert \tau_{4}, 0\vert0) + \\
& + & \sum_{k,n} P(k \vert \tau_{1}, k \vert \tau_{2}, 0\vert0)P(n \vert \tau_{3}, n \vert \tau_{4}, 0\vert0) \, .
\end{eqnarray}
The above summations can be carried out precisely as it was already done in our previous calculations which lead us to (\ref{5}). Consider, for the sake of simplicity, the first sum on the right hand side of (\ref{functionR}). 
It is straightforward to show that
\begin{equation}
\label{var4}
\fl P(k \vert \tau_{1}, k \vert \tau_{2}, 0\vert0)P(k \vert \tau_{3}, k \vert \tau_{4}, 0\vert0) \approx \frac{ \theta_{3}(0; \exp(-B_{12}-B_{34})) }{ (2\pi)^{2} (\tau_{1}\tau_{2}\tau_{3}\tau_{4})^{H} \sqrt{1-g^{2}(\tau_{1},\tau_{2})} \sqrt{1-g^{2}(\tau_{3},\tau_{4})} } \, ,
\end{equation}
where
\begin{equation}
B_{ij} = \frac{1}{1-g^{2}_{ij}} \frac{ |\tau_{i}-\tau_{j}|^{2H} }{2(\tau_{i}\tau_{j})^{2H}} \, ,
\end{equation}
and
\begin{equation}
\theta_{3}(0; \exp(-B_{12}-B_{34})) = \sqrt{\frac{\pi}{B_{12}+B_{34}}} \, .
\end{equation}
Other sums can be tackled in essentially the same way. 

Noticing next that in the limit $f \to 0$ the integrand in (\ref{var3}) stays finite, we find 
\begin{eqnarray} \nonumber
\label{var5}
\fl \textrm{Var}\left(  W_{0}^{(T)} \right) & = & \frac{1}{4T^{2}} \int_{0}^{\infty}\textrm{d}\tau_{1} \cdots \int_{0}^{\infty}\textrm{d}\tau_{4} \, \theta(2T-\tau_{1}-\tau_{2}) \theta(2T-\tau_{3}-\tau_{4}) \times \\
& \times & (2T-\tau_{1}-\tau_{2})^{2}(2T-\tau_{3}-\tau_{4})^{2} R(\tau_{1},\dots,\tau_{4}) \, .
\end{eqnarray}
Similarly to the analysis of the mean value (\ref{mean_zf}), here we perform the following change of the integration variables: $\tau_{j}=Tx_{j}$, $j=1,\dots,4$,  which permits us to cast the variance into the form
\begin{eqnarray} \nonumber
\label{var6}
\textrm{Var}\left(  W_{0}^{(T)} \right) & = & T^{6-3H} \Psi_{2}(H) \, ,
\end{eqnarray}
with $\Psi_{2}(H)$ being a dimensionless function of the Hurst index $H$. 
Combining (\ref{mean_zf}) and (\ref{var6}), we conclude that the probability density characterising the finite-time Wigner-Ville spectrum at zero-frequency has a coefficient of variation $\gamma_{W}\sim T^{2-H/2}$, implying that the distribution broadens with $T$ and the sample-to-sample fluctuations become substantially more important.  However, we are not in position to determine the coefficient of variation for finite $f > 0$ here. In principle,  we expect that similarly to what happens with a super-diffusive fBm, $\gamma_v$ will attain a finite value in the limit $T \to \infty$ (see \cite{Krapf_2019}).  
 This analysis goes beyond the scope of the present work and will be published elsewhere.

\section{Effective couplings in the Ising-like model representation of the Wigner-Ville spectrum}
\label{dis3}

In this last section we represent the Wigner-Ville spectrum $W_f$ as a Hamiltonian of an Ising-like chain of "spin" variables $\sigma_k$, (which prescribe  the directions of the flows), and analyse the form of the effective couplings in this representation. Taking advantage of our previous results, we have that the Laplace-transformed over the observation time $T$ power spectral density admits the following form
\begin{equation}
\label{Ising}
W_f =  \lim_{\lambda \to 0} \sum_{k=-\infty}^{\infty} \sum_{k'=-\infty}^{\infty} J_{k,k'}(\lambda) \sigma_k \sigma_{k'} \,,
\end{equation} 
where the couplings $J_{k,k'}(\lambda)$ are defined by
\begin{eqnarray}
\label{J}
\fl J_{k,k'}(\lambda) & = & \frac{\lambda}{2 \pi f^2} \int^{\infty}_0 \frac{{\rm d}\tau_1}{\tau_1^H} \int^{\infty}_0 \frac{{\rm d}\tau_2}{\tau_2^H \sqrt{1-g^2}} {\rm e}^{- i f (\tau_1 - \tau_2) - \lambda (\tau_1 + \tau_2)/2} \\ \nonumber
\fl & \times & \exp\left( - \frac{g}{2 (1 - g^2)} \left(\frac{k}{\tau_1^H} - \frac{k'}{\tau_2^H}\right)^2  - \frac{1}{2 (1 + g)} \left(\frac{k^2}{\tau_1^{2 H}} + \frac{(k')^2}{\tau_2^{2 H}}\right) \right) \, .
\end{eqnarray}
By noticing that $J_{k,k'}(\lambda)$ are evidently real-valued, even functions of $f$, we can also rewrite (\ref{J}) in the form 
\begin{eqnarray} \nonumber
\label{Jj}
\fl  &J_{k,k'}(\lambda) & =  \frac{\lambda}{2 \pi f^2} \int^{\infty}_0 \frac{{\rm d}\tau_1}{\tau_1^H} \int^{\infty}_0 \frac{{\rm d}\tau_2}{\tau_2^H \sqrt{1-g^2}} \cos\left(f \left(\tau_1 - \tau_2\right)\right) {\rm e}^{ - \lambda (\tau_1 + \tau_2)/2} \times \\ \nonumber
\fl & \times & \exp\left( - \frac{g}{2 (1 - g^2)} \left(\frac{k}{\tau_1^H} - \frac{k'}{\tau_2^H}\right)^2  - \frac{1}{2 (1 + g)} \left(\frac{k^2}{\tau_1^{2 H}} + \frac{(k')^2}{\tau_2^{2 H}}\right) \right) \, , \\
\end{eqnarray}
which is somewhat easier to handle. Still, we are able to determine $J_{k,k'}(\lambda)$ for arbitrary $k$ and $k'$ only for the Brownian motion with $H = 1/2$. For arbitrary $H$, it turns possible only to determine $J_{k,k}(\lambda)$.

\subsection{Particular case $H = 1/2$}
\begin{figure}[htbp]
\mbox{
\includegraphics[width=0.49\textwidth]{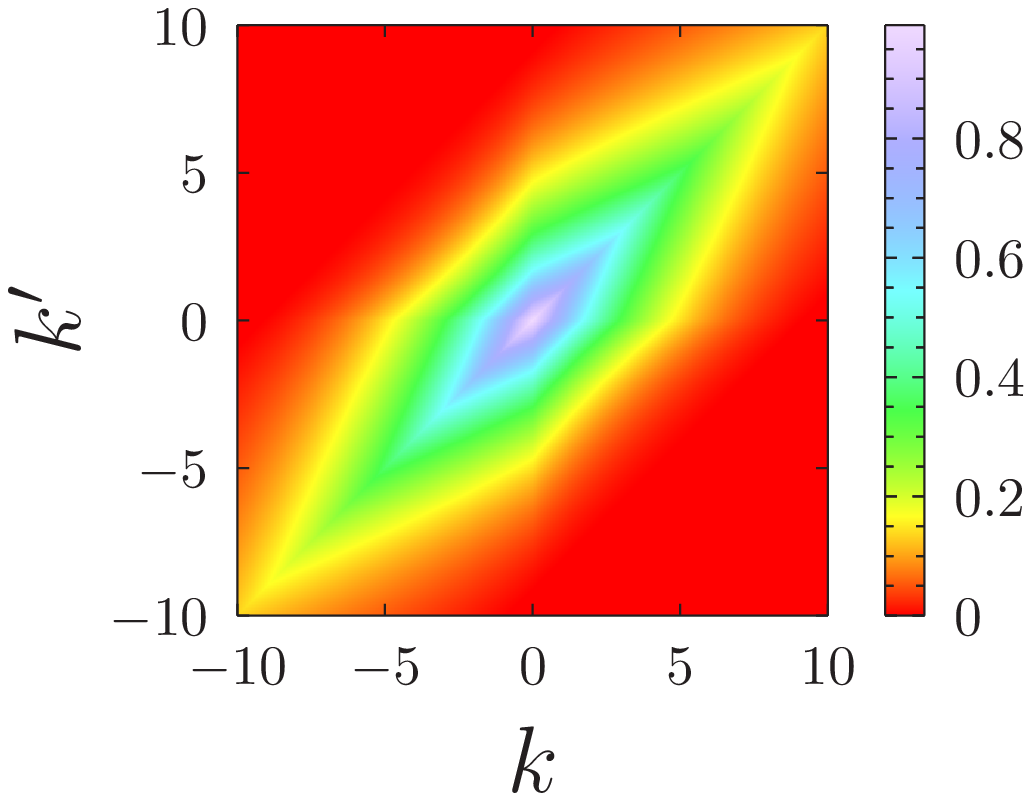}
\includegraphics[width=0.49\textwidth]{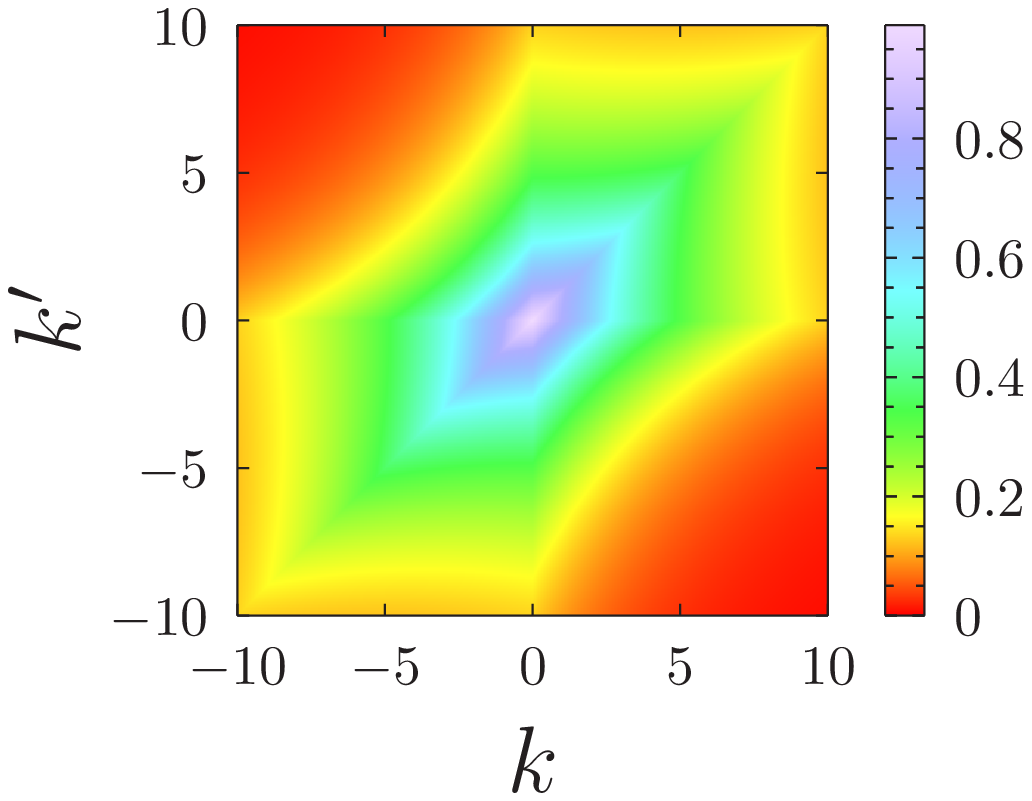}}
\caption{The rescaled couplings $J_{k,k^{\prime}}(\lambda)/J_{0,0}(\lambda)$ for the Brownian motion $X(t)$; see (\ref{BM}). Left panel: $f=5\cdot10^{-3}$, right panel: $f=5\cdot10^{-4}$; in both the figures $\lambda=5\cdot10^{-3}$.}
\label{fig_coupling_Brownian}
\end{figure}

It is expedient to get first some idea on the form of $J_{k,k'}(\lambda)$. This can be done in the special case $H = 1/2$ in which the integrals in (\ref{Jj}) can be performed exactly. After straightforward calculations, we find that for $H = 1/2$ the couplings in (\ref{Ising}) are given explicitly by 
\begin{equation}
\label{BM}
\fl J_{k,k'}(\lambda) = \frac{\sqrt{\lambda}}{\sqrt{8} f^2} \left( {\rm e}^{- |k| \sqrt{2 \lambda}} + {\rm e}^{-|k'| \sqrt{2 \lambda}} \right) \left( \frac{{\rm e}^{-|k - k'| \sqrt{\lambda + 2 i f} }}{\sqrt{\lambda + 2 i f}} + \frac{{\rm e}^{-|k - k'| \sqrt{\lambda - 2 i f} }}{\sqrt{\lambda - 2 i f}}\right) \,,
\end{equation}
which expression holds for any value of the parameters 
$f$, $\lambda$, $k$ and $k'$. Note that $J_{k,k'}(\lambda)$ are real-valued functions and we have chosen the form in (\ref{BM}), which involves the unit imaginary number $i$ just for the sake of compactness. Note, as well, that  $J_{k,k'}(\lambda)$  are oscillatory functions of $|k - k'|$ with a period of oscillations $\sim 1/\sqrt{f}$  in the limit $\lambda \to 0$. In Fig. \ref{fig_coupling_Brownian} we show the couplings $J_{k,k^{\prime}}(\lambda)$ as functions of $k$ and $k^{\prime}$ for a fixed frequency $f$. Quite generally, $J_{k,k^{\prime}}(\lambda)$ reach their maximal values at the origin,  i.e., for $k=k^{\prime}=0$, and vanish for large $k$ and $k^{\prime}$. It is evident from Fig \ref{fig_coupling_Brownian} that the decay of the couplings is spatially anisotropic, with $k=k^{\prime}$ being the direction of a slow decay and $k=-k^{\prime}$ - the direction of a fast decay. Upon decreasing the frequency, the spatial dependence of the couplings acquires further anisotropic structure, as illustrated in the right panel of Fig. \ref{fig_coupling_Brownian}. 

Several points have to be emphasised.\\
 First, we notice that the result in (\ref{BM}) permits us to reproduce (\ref{dis-av}) in the particular case $H = 1/2$. Indeed, noticing that $\langle \sigma_k \sigma_{k'}\rangle = \delta_{k,k'}$,  we have
\begin{eqnarray} \nonumber
\langle W_f\rangle & = & \lim_{\lambda \to 0} \sum_{k=-\infty}^{\infty} J_{k,k}(\lambda) \\
& = & \lim_{\lambda \to 0} \frac{\sqrt{\lambda^2 + \lambda \sqrt{\lambda^2 + 4 f^2}}}{f^2 \sqrt{\lambda^2 + 4 f^2}} \left(1 + 2 \frac{{\rm e}^{-\sqrt{2 \lambda}}}{1 - {\rm e}^{-\sqrt{2 \lambda}}}\right) \equiv \frac{1}{f^{5/2}}
\end{eqnarray}
which is precisely our result in (\ref{BMS}), obtained by taking into account only the leading asymptotic behaviour of the Jacobi theta-function. This means, in turn, that  the arguments used for the derivation of the result in (\ref{BMS}) are correct.\\
Second, we notice that taking the limit $\lambda \to 0$ and performing the summation operations can not be interchanged. Indeed, for any fixed $k$ and $k'$ we have $\lim_{\lambda \to 0}J_{k,k'}(\lambda) = 0$, which signifies, in turn, that for small $\lambda$ the sums in (\ref{Ising}) are dominated by the terms with large $k$ and $k'$. Further, since $J_{k,k'}(\lambda)$ are linked 
to the correlation function of the local occupation times at two different points, they depend simultaneously on both the distances $|k|$ and $|k'|$ from this sites to the origin (the starting point of the trajectory $X(t)$) and also on their relative distance $|k - k'|$. In these dependences, which are simple exponential functions for $H=1/2$, the characteristic decay of $J_{k,k'}(\lambda)$ with $|k-k'|$ depends on both $f$ and $\lambda$ and here we can safely set $\lambda = 0$. On contrary, we have to keep the dependence of the characteristic decay length on $\lambda$ in the terms dependent only on $k$ and $k'$ (the sum of two exponential functions in the first factor in the right hand-side of (\ref{BM}).

\subsection{Arbitrary $H \in (0,1)$.}
We turn next to the general case of an arbitrary $H \in (0,1)$ focussing on the diagonal terms. For $k=k'$, the couplings $J_{k,k}(\lambda)$ in (\ref{Jj}) read
\begin{eqnarray} \nonumber
\label{Jjj}
J_{k,k}(\lambda) & = & \frac{\lambda}{2 \pi f^2} \int^{\infty}_0 \frac{{\rm d}\tau_1}{\tau_1^H} \int^{\infty}_0 \frac{{\rm d}\tau_2}{\tau_2^H \sqrt{1-g^2}} \cos\left(f \left(\tau_1 - \tau_2\right)\right) {\rm e}^{ - \lambda (\tau_1 + \tau_2)/2} \times  \\
& \times & \exp\left( - \frac{|\tau_1 - \tau_2|^{2 H}}{2 \left(1 - g^2\right) \tau_1^{2 H} \tau_2^{2 H}} \, k^2 \right) \, .
\end{eqnarray}
It is convenient next to change the integration variables $\tau_2 = \tau_1 \xi$ and $\tau_1 = \phi/\lambda$ and rewrite $J_{k,k}(\lambda)$ in (\ref{Jjj}) as
\begin{eqnarray} \nonumber
\label{J4}
J_{k,k}(\lambda) & = & \frac{\lambda^{2 H -1}}{2 \pi f^2} \int^{\infty}_0 \phi^{1 - 2 H} {\rm d}\phi \int^{\infty}_0 \frac{{\rm d}\xi}{\xi^H \sqrt{1-\phi^2(1,\xi)}} \cos\left(\frac{f}{\lambda} \phi \left(1 - \xi\right)\right) \times \\
& \times &  {\rm e}^{ - \phi (1 + \xi)/2} \exp\left( - \frac{|1 - \xi|^{2 H}}{2 \left(1 - \phi^2(1,\xi)\right) \xi^{2 H} \phi^{2 H}} \, \left(k \lambda^H\right)^2 \right) \, ,
\end{eqnarray}
where $\phi(1,\xi)$ is defined in (\ref{phi}). Inspecting the integral over $\textrm{d}\xi$, we note that the cosine term oscillates heavily when $f$ is kept fixed and $\lambda \to 0$, which means that in this limit the integral over $d\xi$ is concentrated in the vicinity of $\xi = 1$, i.e., for $\tau_2 \approx \tau_1$. Expanding then $\phi(1,\xi)$ in the vicinity of $\xi = 1$, 
\begin{equation}
\phi(1,\xi) = 1 - \frac{1}{2} |1-\xi|^{2 H} + O\left((1 - \xi)^2\right) \,,
\end{equation}
as well as other $\xi$-dependent functions in the kernel and taking into account only the leading in this limit terms, we find that the integral in (\ref{J4}) can be approximately rewritten as
\begin{equation}
\label{J5}\fl
J_{k,k}(\lambda) \approx \frac{\lambda^{2 H -1}}{2 \pi f^2} \int^{\infty}_0 \phi^{1 - 2 H} {\rm e}^{- \phi} \exp\left( - \frac{\left(k \lambda^H\right)^2}{2 \phi^{2 H}} \right) {\rm d}\phi \, \int^{\infty}_{-\infty} \frac{{\rm d}\xi}{|1-\xi|^H}  \cos\left(\frac{f}{\lambda} \phi (1 - \xi)\right) \, .
\end{equation}
Performing then the integral over $d \xi$, we find
\begin{equation}
\label{J5}\fl
J_{k,k}(\lambda) \approx \frac{\Gamma(1-H) \sin\left(\frac{\pi H}{2}\right)}{\pi} \frac{\lambda^H}{f^{3 - H}} \int^{\infty}_0 \frac{d\phi}{\phi^H} e^{-\phi} \exp\left(- \frac{\left(k \lambda^H\right)^2}{2 \phi^{2 H}}\right) \, ,
\end{equation}
which form often arises in the analysis of various physical problems via the optimal fluctuation method (see, e.g., Ref. \cite{trap1,trap}). On the other hand, the integral in (\ref{J5}) defines the so-called special  
Kr\"atzel function (see, e.g., Ref. \cite{tru})
\begin{equation}
\label{ }
Z_{\rho}^{\nu}(x) = \int_{0}^{\infty}\textrm{d}z \, z^{\nu-1} \exp\left( -z^{\rho} - x/z \right) \, ,
\end{equation}
and therefore, the diagonal couplings can be formally represented as
\begin{equation}
\label{J7}
J_{k,k}(\lambda) \approx \frac{\Gamma(1-H) \sin\left(\frac{\pi H}{2}\right)}{2 H \pi} \frac{\lambda^H}{f^{3 - H}} Z^{(1-H)/(2 H)}_{1/(2 H)}\left( \left(k \lambda^H\right)^2/2 \right) \, .
\end{equation}
Before we proceed to the analysis of the asymptotic behaviour of $J_{k,k}(\lambda)$ in (\ref{J7}), it seems expedient to consider first a particular case of Brownian motion ($H=1/2$). Here, we have
\begin{equation}
\label{ }
Z^{1/2}_{1}(x) = \sqrt{\pi} {\rm e}^{-2 \sqrt{x}} \, ,
\end{equation}
such that
\begin{equation}
\label{001}
J_{k,k}(\lambda) =\frac{\sqrt{\lambda}}{\sqrt{2} f^{5/2}} \, {\rm e}^{- |k| \sqrt{2 \lambda}} \, ,
\end{equation}
which expression agrees perfectly well 
with our previous result in (\ref{BM}) with $k$ set equal to $k^{\prime}$. Further on, summing the expression in the second line in (\ref{J5}) over all $k$, and taking an appropriate limit in the resulting Jacobi theta-function, we recover exactly our result in (\ref{dis-av}) for the disorder-averaged Wigner-Ville spectrum.

\begin{figure}[htbp]
\begin{center}
\centerline{\includegraphics[width = 120mm]{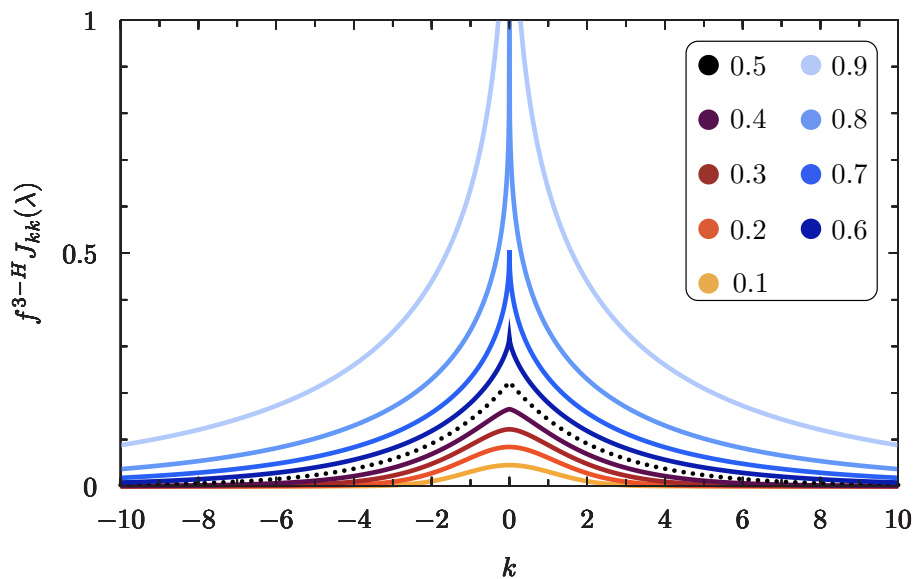}}
\caption{The rescaled diagonal couplings $J_{k,k}(\lambda) f^{3-H}$ given by (\ref{J7}) as functions of $k$ for the values of $H$ shown in the inset. In this figure $\lambda=0.1$; different values don't change the plot in a qualitative way.}
\label{Fig_couplings_fBm}
\end{center}
\end{figure}

The $k$-dependence of the diagonal couplings given in (\ref{J7}) is plotted in Fig. \ref{Fig_couplings_fBm}. The case of the Brownian motion, corresponding to (\ref{BM}) with $k=k^{\prime}$, is depicted with a dotted line. We observe that upon increasing the Hurst index, the diagonal coupling increases monotonically for any $k$, while for any fixed $H$ the large-$|k|$ behavior indicates the stretched-exponential decay predicted by the asymptotic result (\ref{stretchedexponential}).

Lastly, we consider the asymptotic behaviour of  $J_{k,k}$ in (\ref{J7}) in the limits $\equiv |k| \lambda^H \ll 1$ and $|k| \lambda^H \to \infty$. For the sake of notational convenience, we rescale $k$ by defining $\kappa=k \lambda^{H}/\sqrt{2}$.
\begin{itemize}
\item Small-$\kappa$ asymptotic behaviour. The leading and the first sub-leading terms in the expansion of the Kr\"atzel function obey \cite{tru}
\begin{equation}
\fl Z^{(1-H)/(2 H)}_{1/(2 H)}\left( \kappa^{2} \right) \sim 2 H \Gamma\left(1 - H\right) + \Gamma\left(\frac{H - 1}{2 H}\right) |\kappa| ^{(1-H)/H} \,,
\end{equation}
so that $J_{k,k}(\lambda)$ in (\ref{J7}) follows
\begin{equation}
\fl J_{k,k}(\lambda) \approx \frac{\Gamma^2\left(1 - H\right) \sin\left(\frac{\pi H}{2}\right)}{\pi} \frac{\lambda^H}{f^{3 - H}} \left(1 + \frac{\Gamma\left(\frac{H - 1}{2 H}\right)}{2 H \Gamma\left(1 - H\right)} |\kappa|^{(1-H)/H}\right) \, .
\end{equation}
\item Large-$\kappa$ asymptotic behaviour. Recalling the properties of the Kr\"atzel function summarised in Ref. \cite{tru}, we find
\begin{eqnarray} \nonumber
\fl Z^{(1-H)/(2 H)}_{1/(2 H)}\left( \kappa^{2} \right) & \sim & 2 \left(\frac{\pi H}{2 H + 1}\right)^{1/2} \left(2 H\right)^{1/(2 H+1)} |\kappa|^{(1 -  2 H)/(1 + 2 H)} \\
& \times & \exp\left( - \frac{(2 H + 1)}{\left(2 H\right)^{2 H/(2 H + 1)}} |\kappa|^{2/(2 H + 1)}\right) \,.
\end{eqnarray}
Therefore, the couplings $J_{k,k}(\lambda)$ in (\ref{J7}) exhibit a stretched-exponential dependence on both $k$ and $\lambda$
\begin{eqnarray} \nonumber
\label{stretchedexponential}
\fl J_{k,k}(\lambda) \approx \frac{\Gamma\left(1 - H\right) \sin\left(\frac{\pi H}{2}\right)}{\sqrt{\pi H (2 H + 1)}} \left(2 H\right)^{1/(2 H+1)} |\kappa|^{(1 -  2 H)/(1 + 2 H)}    \frac{\lambda^{H}}{f^{3 - H}} \times \\
\times \exp\left( - \frac{(2 H + 1)}{\left(2 H\right)^{2 H/(2 H + 1)}} |\kappa|^{2/(2 H + 1)}\right) \, .
\end{eqnarray}
In the particular case $H=1/2$, the decay with $k$ in the latter expression becomes purely exponential. For super-diffusive fBm process $X(t)$, the decay
of  $J_{k,k}(\lambda)$ with $k$ is slower than exponential, i.e., $J_{k,k}(\lambda) \sim \exp\left(- |k|^z\right)$ with $z = 2/(2H+1)< 1$. On contrary, for sub-diffusive processes $X(t)$, $J_{k,k}(\lambda)$ vanishes faster than an exponential,  $J_{k,k}(\lambda) \sim \exp\left(- |k|^z\right)$ with $z = 2/(2H+1) > 1$. In particular, for $H = 1/4$ (the case of a tagged bead in an infinite Rouse polymer), one finds
$J_{k,k}(\lambda) \sim \exp\left(- |k|^{4/3}\right)$.
\end{itemize}

\section{Conclusions}
\label{dis4}
To recap, we have studied here a model of anomalous diffusion of a tracer particle 
in presence of
stratified layers of quenched random flows. The particle undergoes a fractional Brownian (fBm) motion with the Hurst index $H \in (0,1)$ in the direction perpendicular to the flows (along the $X$-axis) and  is passively advected in random directions (with no global bias) by the flows along the $Y$-axis. 
Averaging the squared displacement $Y^2(t)$ along the flows over the flow realisations, as well as over the fBm-trajectories, we obtained the mean-squared displacement along the flow direction and showed that for large times it grows in proportion to $t^{2-H}$, which law defines a family of super-diffusive processes. In order to quantify the sample-to-sample fluctuations of the displacement along the random flow, we computed the coefficient of variation of the probability distribution function of the displacement. We showed that such fluctuations are typically not very important for $H\gtrsim0.22$ but may become important for sufficiently small values of $H$. 

Next,  we  introduced the Wigner-Ville power spectral density (see (\ref{spec2}) and (\ref{spec3})) of the random process $Y(t)$ and derived an exact result showing  that the disorder-averaged Wigner-Ville spectrum $<W_f>$ scales with the frequency as $f^{H-3}$, 
which defines a family of rather unusual super-diffusive spectra. To estimate the sample-to-sample fluctuations of the spectrum,  we examined the limit of zero frequency of the Wigner-Ville spectrum averaged over a finite time interval of duration $T$. We found that the ratio of the standard deviation of the spectrum and of the mean value diverges with $T$, which implies that the distribution of the spectrum with respect to different realisations of disorder progressively broadens with the observation time. 

Lastly, representing the Wigner-Ville spectrum of the process $Y(t)$ for a given realisation of flows as an Ising-like model
of "spins"  $\sigma_k$, which define the local direction of the flows along the $X$-axis, we have analytically determined the coupling terms in this model, in the Laplace domain.  In particular, we showed that the latter exhibit a non-trivial dependence on the distance from the origin and the Laplace parameter.

In general, our analysis reveals that sample-to-sample fluctuations are important, both for the dynamical characteristics and the spectral behaviour. This issue is beyond the scope of the current work and will be examined elsewhere.



\end{document}